\documentstyle{elsart}
\begin{document}
\begin{frontmatter} 
\title{The mixed quark-gluon condensate from an effective 
quark-quark interaction}
%\thanks{supported by NSF grant \# PHY-9319641.}
\author{T. Meissner\thanksref{mei}}
\address{Department of Physics, Carnegie Mellon University, 
Pittsburgh, PA 15213, U.S.A.}
\thanks[mei]{email: meissner@yukawa.phys.cmu.edu}
\begin{abstract}
We exhibit the method for obtaining non perturbative quark and 
gluonic vacuum condensates from a model truncation of QCD.
The truncation allows for a phenomenological description of the
quark-quark interaction in a framework which maintains all 
global symmetries of QCD and allows an $\frac{1}{N_C}$ expansion. 
Within this approach the functional integration over the gluon fields
can be performed and therefore
any gluonic vacuum observable can be expressed
in terms of a quark operator and the gluon propagator.  
As a special case we calculate the mixed quark gluon condensate
$ g_s \langle {\bar q} \, G_{\mu\nu} \sigma^{\mu\nu} \, q \rangle $.
We investigate how the value depends on the form of the 
model quark-quark interaction. 
A comparison with the results of quenched 
lattice QCD, the instanton liquid model
and QCD sum rules is drawn.
\end{abstract}
\end{frontmatter}
The existence of finite vacuum condensates such as 
the quark condensate $\langle {\bar q} q \rangle$,
the mixed quark gluon condensate $g_s \langle {\bar q} \, G_{\mu\nu} \sigma^{\mu\nu} q \rangle$, 
the gluon condensate $\alpha_s \langle G_{\mu\nu}G^{\mu\nu} \rangle$ or
the four quark condensate $\langle  {\bar q} \Gamma q {\bar q} \Gamma q \rangle$
reflects in a direct way the non perturbative
structure of the QCD vacuum.
Their determination within a certain
approach provides therefore important information about its ability to 
describe sufficiently the physics of strong interaction at low and intermediate
energies.
One major field of application of these condensates are QCD sum rules
\cite{SVZ,RRY,Nar2}, which are based on the operator product expansion (OPE).  
Hereby the vacuum condensates 
serve as the basic input parameters 
in which all the non perturbative effects are incorporated whereas all the 
short distance effects are treated within the standard perturbative diagram
technique. 
The knowledge of the vacuum values at least of the lower dimensional quark and
gluonic operators
is therefore essential for the 
applicability of the QCD sum rule method. 

In this paper we want to focus on the mixed condensate 
$g_s \langle {\bar q} \, G_{\mu\nu} \sigma^{\mu\nu} q \rangle$  
of dimension five, which is
among the least well determined ones.
Previous studies of this condensate
include QCD sum rules themselves where it was treated as fit parameter
in the analysis of heavy-light quark system spectra \cite{Nar2,Nar1}, quenched
lattice QCD \cite{Schierholz} and, very recently, the 
instanton liquid model \cite{weiss}. 

It is the aim of this letter to consider vacuum condensates in general and in particular the
mixed condensate 
$g_s \langle {\bar q} \, G_{\mu\nu} \sigma^{\mu\nu} q \rangle$ in the framework of 
a truncation of QCD which is based on an effective quark-quark interaction.  
For this let us consider the QCD partition function for massless quarks in Euclidean space 
\begin{equation}
{\cal Z}_{\mathrm{QCD}} \equiv \int {\cal D} q {\cal D} {\bar q} {\cal D} A \e^{- S_{\mathrm{QCD}}[q,{\bar q},A]}
\label{zqcd}
\end{equation}
with the Euclidean action
\begin{equation}
S_{\mathrm{QCD}} [q,{\bar q},A] = \int d x \left \{ {\bar q} ( \FMSlash{\partial} -i g_s \FMSlash{A} ) q 
+\quart G_{\mu\nu}^a G_{\mu\nu}^a \right \}
\label{sqcd}
\end{equation}
where
\begin{equation}
{A_\mu} = A_{\mu}^a \frac{\lambda^a}{2} \; , \; 
G_{\mu\nu} = G_{\mu\nu}^a \frac{\lambda^a}{2}  \; , \; 
G_{\mu\nu}^a = \partial_\mu A_{\nu}^a - \partial_\nu A_{\mu}^a + g_s f^{abc} A_{\mu}^b A_{\nu}^c
\quad .
\label{gdef}
\end{equation}
Eq.(\ref{zqcd}) can be rewritten as
\begin{equation}
{\cal Z}_{\mathrm{QCD}} \equiv \int D q D {\bar q} \e^{- {\bar q }\FMSlash{\partial} q }
\e^{{\cal W} [ i g_s {\bar q} \gamma_\mu \frac{\lambda^a}{2}  q ] } 
\label{zqcd2}
\end{equation}
where
\begin{equation}
\e^{{\cal W} [j]} \equiv \int {\cal D} A \e^{  \int 
\left (-  \quart G_{\mu\nu}^a G_{\mu\nu}^a  + j_\mu ^a A_{\mu}^a 
\right ) }
\quad .
\label{wdef}
\end{equation}
The functional ${\cal W} [j]$ can be formally expanded in the current $j_\mu ^a$ which leads to an
expansion in terms of gluon $n$-point functions,
\begin{eqnarray}
{\cal W} [j] 
\equiv 
&\half& 
\int 
d x_1 \,
d x_2 \,
j_{\mu_1}^{a_1} (x_1) \, 
D_{{\mu_1}{\mu_2}}^{a_1 a_2} (x_1, x_2) \, 
j_{\mu_2}^{a_2} (x_2) \,
\, +  \nonumber \\
&{\frac{1}{3!}}& 
\int 
D_{{\mu_1}{\mu_2}{\mu_3} } ^{a_1 a_2 a_3}  \,
j_{\mu_1}^{a_1} \, j_{\mu_2}^{a_2} \, j_{\mu_3}^{a_3} \,
\, + \, \dots
\quad .
\label{expansion}
\end{eqnarray}
The first nontrivial contribution arises from the gluon 2-point function
\begin{equation}
D_{\mu\nu}^{ab} (x,y) =  
D_{\mu\nu}^{ab} (x-y) =
\int {\cal D} A A_\mu^a (x) A_\nu^b (y) \e^{-S_{\mathrm{QCD}}[0,0,A]}
\quad . 
\label{gluonprop}
\end{equation}

It should be noted that in this definition of $D_{\mu\nu}$ does not 
include quark loops.

Our model truncation consists in truncating the series
(\ref{expansion}) after $n=2$, i.e. we include only the gluon 2-point function.
This defines an {\em effective} model based on the bilocal quark-quark interaction kernel
$D_{\mu \nu }^{ab} (x-y)$.
This model truncation respects all global symmetries of QCD. 
It has been used extensively in the literature and is known as the 
{\em global color model} (GCM)
\cite{cahill,gsm,kekez,robertswilliams,tandy}. 
The main features of QCD, which are lost by this truncation are
local SU(3) gauge invariance and renormalizability.

The partition function of this truncation is given by
\begin{equation}
{\cal Z}_{\mathrm{GCM}}  = 
\int {\cal D} q {\cal D} {\bar q}
\e^{- \left  \{  \int {\bar q }\FMSlash{\partial} q
+ \frac{g_s^2}{2}  
\int 
{d x d y
\left [ {\bar q} (x) \gamma_\mu \frac{\lambda^a}{2} q(x) \right ]
D_{\mu \nu }^{ab}(x-y)
\left [ {\bar q} (y) \gamma_\nu \frac{\lambda^b}{2} q(y) \right ]} \,
\right \} } 
\label{zgcma}
\end{equation} 
or
\begin{equation}
{\cal Z}_{\mathrm{GCM}} 
= 
\int {\cal D} q {\cal D} {\bar q}  {\cal D} A
\e^{- \left \{
\int {\bar q} \left ( \FMSlash{\partial} - i g_s  \FMSlash{A} \right ) q
+ \int \int 
\half A D^{-1} A 
\right \} }
\quad  .
\label{zgcmb}
\end{equation}
Eqs. (\ref{zgcma}) and (\ref{zgcmb}) are connected by the functional integration rule
\begin{equation}
\e^{W[j]} = \e^{ \half j D j} = \int {\cal D} A \e^{ \left [ - \half A D^{-1} A +  j A \right ] }
\label{ii1}
\end{equation}
where an obvious short hand notation for the integrations has been used.

The gluon 2-point function $D_{\mu \nu }^{ab} (x-y)$ is treated as the model input parameter, which,
as we will discuss later, is chosen to reproduce certain aspects of low energy hadronic physics. 
For convenience we will use the Feynman like gauge
$D_{\mu \nu }(x-y) = \delta_{\mu\nu} \delta^{ab} D(x-y)$ from now on.

The next step is to apply the standard bosonization procedure , 
which consists in rewriting the partition function in terms
of bilocal meson-like integration variables and expanding about the the classical vacuum, i.e. 
the saddle point of the action \cite{cahill,gsm,frankmeissner1}.

The resulting expression for the partition function in terms of the 
bilocal field integration is
${\cal Z}_{\mathrm{GCM}} =   
\int {\cal D} 
{\cal B}
e^{-{\cal S}[{\cal B}]}$ 
where the action is given by
\begin{equation}
{\cal S}[{\cal B}]=
-
{\mathrm{TrLn}}
\left[ S^{-1}\right] +\int d x
d  y 
\frac{{\cal B}^{\theta 
}(x,y){\cal B}^{\theta }(y,x)}
{2 {g_s}^2 D(x-y)}
\label{bilocal}
\end{equation}
and the quark inverse Green's function, $S^{-1}$, is defined as 
\begin{equation} 
S^{-1}(x,y)
\equiv  \FMSlash{\partial}_x
\delta (x - y)
+\Lambda ^{\theta }{\cal B}^{\theta }(x,y)
\quad  .
\label{gf}
\end{equation}  
The quantity $\Lambda ^{\theta }$
arises from Fierz reordering the current-current interaction in
(\ref{zgcma})  
\begin{equation}
{\Lambda^{\theta}}_{ji}  
{\Lambda^{\theta}}_{lk}  
=
\left ( \gamma_\mu \frac{\lambda^a}{2} \right )_{jk}
\left ( \gamma_\mu \frac{\lambda^a}{2} \right )_{li}
\quad   .
\label{fierz2}
\end{equation}
It is the direct product of Dirac, flavor $SU(3)$  and color matrices
\begin{eqnarray}
\Lambda ^{\theta }
&=&
\frac{1}{2}\left( {\bf 1}_{D},i\gamma 
_{5},\frac{i}{\sqrt{2}}\gamma _{\nu },\frac{i}{\sqrt{2}}\gamma _{\nu 
}\gamma _{5}\right) 
\\ \nonumber
&\otimes&
 \left( \frac{1}{\sqrt{3}}{\bf 
1}_{F},\frac{1}{\sqrt{2}}\lambda _F^a\right) 
\\ \nonumber
&\otimes&
\left( \frac{4}{3}{\bf 
1}_{C},\frac{i}{\sqrt{3}}\lambda _C^{a}\right) 
\quad    .
\label{lamdadef}
\end{eqnarray}

The saddle-point of the action is defined as
$\frac{\delta S}{\delta {\cal B}} \vert_{{\cal B}_0}$ and 
is given by
\begin{equation}
{\cal B}^{\theta }_0 (x-y)
=
g_s^2
D(x-y) {\mathrm{tr}}
\left[ \Lambda ^{\theta } S_0 (x-y) \right ] 
\label{spoint}
\end{equation}
where $S_0$ stands for $S[{\cal B}^{\theta }_0]$.  
These configurations 
provide self-energy dressing of the quarks with ``rainbow'' gluons 
through the definition 
$\Sigma (p)\equiv \Lambda ^{\theta }{\cal B}^{\theta }_0(p)=
i  \FMSlash{p}
\left[ A(p^2)-1\right] +B(p^2)$. 
The self energy functions $A$ and $B$
are determined by the ``rainbow'' Dyson-Schwinger equation
\begin{eqnarray}
\left[ A(p^{2})-1\right] p^{2} &=&
\frac{8}{3} \int \frac{d^4 q}
{(2\pi 
)^{4}} g_s^{2}
D(p-q)
\frac{A(q^{2})q\cdot p}{q^{2}A^{2}(q^{2})+B^{2}(q^{2})}
\\ \nonumber
B(p^{2})
&=& 
\frac{16}{3}\int \frac{d^4 q}
{(2\pi 
)^{4}}
g_s^{2} D(p-q)
\frac{B(q^{2})}{q^{2}A^{2}(q^{2})+B^{2}(q^{2})} 
\quad  .
\label{dysonschwinger}
\end{eqnarray}
This dressing comprises the 
notion of ``constituent'' quarks by providing a mass
$M(p^2)=B(p^2)/A(p^2)$, reflecting a vacuum configuration with
dynamically broken chiral symmetry.

We will calculate the vacuum condensates from the above saddle-point expansion, that is,
we will work at the mean field level.
This is consistent with the 
lowest order $\frac{1}{N_C}$ in the quark fields for a given model gluon 2-point function. 
It should be noted, however, that the gluon 2-point function $D$
itself contains all powers of $\frac{1}{N_C}$.

The mesonic modes can be obtained within in the standard RPA technique by
considering small fluctuations around the saddle point configuration and 
solving for the eigenmodes of the arising quadratic kernels.
This leads then automatically to the
ladder Bethe Salpeter equations for the ${\bar q} q$ bound states. 
Mesonic properties have been extensively studied in refs.\cite{cahill,gsm}.
In ref.\cite{frankmeissner1} a detailed investigation of the low energy
sector was performed
by deriving the general form of the effective chiral action for the SU(3) Goldstone bosons
and determining $f_\pi$ and most of the chiral low energy coefficients (Gasser-Leutwyler coefficients) $L_i$,
which, in turn, determines the physics of the $\pi$, $K$ and $\eta$ mesons at low energies \cite{chipth}.
This was done for various forms of the model gluon 2-point function $D(q^2)$
all of them reproducing $f_\pi$. 
In the present approach we are following this spirit by choosing gluon 2-point functions 
which have this feature and
moreover reproduce values for the chiral coefficient $L_i \, , \, i=1,2,3,5,8$ which are compatible
with the phenomenological ones \cite{chipth}.

It is now rather straightforward to calculate the vacuum expectation value of any
quark operator of the form 
\begin{equation}
{\cal Q}_n \; \equiv \;
\left ( {\bar q}_{j_1} \Lambda_{j_1 i_1}^{(1)} q_{i_1} \right )
\left ( {\bar q}_{j_2} \Lambda_{j_2 i_2}^{(2)} q_{i_2} \right )
\dots
\left ( {\bar q}_{j_n} \Lambda_{j_n i_n}^{(n)} q_{i_n} \right )
\label{qoperator}
\end{equation}
in the mean field vacuum.
Here the $\Lambda^{(i)}$ stands for an operator in Dirac, flavor or color space.  
This can be done by defining the functional
\begin{equation}
{\cal G} [\eta,{\bar \eta}] 
\; \equiv \;
\frac{ 
\int {\cal D} q {\cal D} {\bar q}
\e^{ - \sum\limits_{ij} {\bar q}_i (S_0 ^{-1})_{ij} q_j 
+ \sum\limits_i ( {\bar \eta}_i q_i + {\bar q}_i \eta_i ) }    }
{
\int {\cal D} q {\cal D} {\bar q}
\e^{ - \sum\limits_{ij} {\bar q}_i (S_0 ^{-1})_{ij} q_j } } 
\; \equiv \;
\e^{\, \sum\limits_{ij}  {\bar \eta}_i (S_0 )_{ij} \eta_j }
\; ,
\label{gg}
\end{equation}
taking the appropriate number of derivatives with respect to external sources $\eta_i$ and ${\bar\eta}_j$
and putting $\eta_i =0$ and ${\bar\eta}_j =0$ \cite{negele}.
This gives
\begin{equation}
\left \langle
: {\cal Q}_n :
 \right \rangle
\; = \;
(-)^n \sum\limits_\Pi 
(-)^{\Pi}
\left \{
\Lambda_{j_1 i_1}^{(1)}
\dots
\Lambda_{j_n i_n}^{(n)}
 \;
(S_0)_{i_1 j_{\pi(1)} }
\dots
(S_0)_{i_n j_{\pi(n)} }
\right \}
\label{qcon}
\end{equation} 
where $\Pi$ stands for a permutation of the $n$ indices.
In particular we obtain for the quark condensate $\langle {\bar q} q\rangle $
\begin{equation}
\langle  {\bar q} q \rangle_\mu
= (-) {\mathrm{tr}}_{\gamma C}
 \{ S_0 (x,x) \}_{x=0} =
(-) \left (  \frac{ N_C}{16\pi^2} \right ) \left \{
4 \int\limits_0^\mu 
d s s \frac{B(s)}{X(s)} \right \}
\label{qbarq}
\end{equation}
where $X(s) \equiv s A(s)^2 + B(s)^2 $.
$\mu$ is the renormalization scale which we chose
to be $1 \mathrm{GeV}^2$.
As indicated the trace in eq.(\ref{qbarq}) is to be taken in Dirac and color space, whereas the flavor trace
has been separated out.

Another important consequence from (\ref{qcon}) is
the fact that the four quark operators factorize in the mean field vacuum. 
For instance in case of $\Gamma^{(1)} = \Gamma^{(2)} = \gamma_\mu \frac{\lambda_C  ^a}{2} $
one finds from (\ref{qcon}) and after reapplying the Fierz transformation (\ref{fierz2})
\begin{equation}
\left \langle : 
\left ( {\bar q} \gamma_\mu \frac{\lambda_C ^a}{2} q \right )
\left ( {\bar q} \gamma_\mu \frac{\lambda_C ^a}{2} q \right )
: \right \rangle 
= (-) \quart \, \frac{16}{9}  \langle  {\bar q} q \rangle^2
\label{factorization}
\end{equation}
i.e. our approach is consistent with the vacuum saturation assumption of ref.\cite{SVZ}.
The fact, that the four quark condensate factorizes due to eq.(\ref{factorization}) is
independent of the model gluon 2-point function which is used.
That means that there are
no four quark correlations at the mean field level, which is of course expected in the large $N_C$ limit.

Let us now turn to gluonic observables.
Because the functional integration over the $A$ field in (\ref{zgcmb}) is quadratic for a given
quark-quark interaction $D$ we can perform the integration over any number of gluon fields 
analytically.
We obtain as a 
generalization of (\ref{ii1}) using the same short hand notation 
\begin{eqnarray}
\int {\cal D} A \e^{ - \half A D^{-1} A +  j A }  &\equiv&  \; \e^{\half j D j} \nonumber \\
\int {\cal D} A A \e^{ - \half A D^{-1} A +  j A }  &\equiv&  \; (j D) (y_1) \, 
\e^{\half j D j} \nonumber \\
\int {\cal D} A A^2 \e^{ - \half A D^{-1} A +  j A }  &\equiv&  \; 
\left  [ D(y_1, y_2) \, + (j D)^2  \right  ] 
\,\, \e^{\half j D j} \nonumber \\
\dots
\quad .
\label{jj2}
\end{eqnarray}
This means that the gluon vacuum average renders effectively a 
quark color current ${\bar q} \gamma_\mu \frac{\lambda_C  ^a}{2} q $
together with the gluon 2-point function $D$.
The integration over the quark operators is then performed in the mean field vacuum as described above.
The remaining expression contains the gluon 2-point function $D$ as well as the quark propagator $S_0$,
which, in turn is given by the self energy functions $A$ and $B$ through the 
Dyson-Schwinger equation (\ref{dysonschwinger}).
The integrations are most conveniently performed in momentum space.  
This way we can in principle obtain the vacuum expectation value for any gluonic and combined
quark-gluonic operator.
It should be stated, however, that the number of terms produced by (\ref{jj2})
will increase rapidly with the number of gluon fields.
For instance for the gluon condensate $\langle G^2 \rangle$, which contains an $A^4$ integration the calculation
gets already rather involved. 

%\begin{samepage}
In case of the mixed condensate $g_s \langle {\bar q} \, G_{\mu\nu} \sigma^{\mu\nu} q \rangle$
we have to integrate only over powers $A^1$ and $A^2$ and therefore
the procedure is still feasible.
Applying the method described above we find
\begin{eqnarray}
&{g_s}& \langle {\bar q} (x) \, G_{\mu\nu} (x) \sigma^{\mu\nu} q (x) \rangle \; = \nonumber \\ 
(-2 i) &{N_C}& \int d z \, \left [ \partial_\mu^{(x)} g_s^2 D(z-x) \right ]
{\mathrm{tr}}_\gamma  [ S_0 (z,x) \sigma_{\mu\nu} S_0 (x,z) \gamma_\nu ] \; + \; \nonumber \\
(+4 i) &{N_C}&  \int d z_1 d z_2  \, g_s^2 D(z_1 -x)  \, g_s^2 D(z_2 -x) 
\nonumber \\ 
&\cdot&
{\mathrm{tr}}_\gamma [ S_0 (z_2,x) \sigma_{\mu\nu} S_0 (x,z_1) \gamma_\mu S_0 (z_1,z_2)  \gamma_\nu ] \quad .
\label{mixed1}
\end{eqnarray}
%\end{samepage}

In order to calculate this expression we can make explicit use of the Dyson-Schwinger equation 
(\ref{spoint}) 
which determines the mean field vacuum configuration
and which can be cast into the form
\begin{equation}
\frac{4}{3} g_s^2 D(x-y) S_0 (x-y) \, = \,
\quart \quart \, {\mathrm{tr}}_\gamma [ \Sigma (x-y) ] 
\, - \,
\half \quart \gamma_\nu  {\mathrm{tr}}_\gamma [ \Sigma (x-y) \gamma_\nu ] 
\, \, .
\label{dg0}
\end{equation}
This eliminates the integration over the
gluon 2-point function in favor of the quark self energy
and therefore strongly simplifies the evaluation of (\ref{mixed1}).
We obtain as final result for the mixed condensate in Minkowski space 
\begin{eqnarray}
g_s \langle  {\bar q} G_{\mu\nu} \sigma^{\mu\nu} q \rangle_\mu
\; = \; 
&(-)& \left (  \frac{N_C}{16\pi^2} \right ) 
\left \{
\frac{27}{4} 
\int\limits_0^\mu 
d s s 
\frac{B}{X} 
\left [ 2 A (A -1)s + B^2 \right ] 
\right \}
\, + \nonumber \\
&(-)& \left (  \frac{N_C}{16\pi^2} \right )
\left \{
12 \int\limits_0^\mu 
d s {s^2} 
\frac{B}{X} 
(2 - A) 
\right \}
\, .
\label{mixed2}
\end{eqnarray}

In tables \ref{tab1},\ref{tab2} and \ref{tab3} we display the result for 
$\langle  {\bar q} q \rangle$ and $ g_s \langle  {\bar q} \sigma G q \rangle$
for three different model gluon 2-point functions, which are parametrized 
by a ``width'' parameter $\Delta$ and a ``strength'' parameter $\chi$
\cite{frankmeissner1}.
As stated above in all cases we fix the pion decay constant 
in the chiral limit to $f_\pi = 87 \mathrm{MeV}$ \cite{chipth}.
The renormalization point is in all cases $\mu = 1 \mathrm{GeV}^2$. 
We also show in all three cases the values obtained for the 
chiral low energy coefficients
$L_i,\, i=1,3,5,8$.
As we can see the values we obtain for the $L_i$ after fixing $f_\pi$ are compatible
with the phenomenological ones \cite{chipth}: \\
$L_1 = 0.7 \pm 0.5$,
$L_3 = -3.6 \pm 1.3$, $L_5 = 1.4 \pm 0.5$, $L_8 = 0.9 \pm 0.3$.

Finally in table \ref{tab4} we compare with the values which other nonperturbative approaches find
for $\langle  {\bar q} q \rangle$ and 
$ g_s \langle  {\bar q} G_{\mu\nu} \sigma^{\mu\nu} q \rangle$:
QCD sum rules \cite{Nar2}, quenched lattice QCD \cite{Schierholz}
and the instanton liquid model \cite{weiss}.
As we can see our results for $ g_s \langle  {\bar q} G_{\mu\nu} \sigma^{\mu\nu} q \rangle$
are compatible with the range obtained within these methods,
whereas the quark condensate  $\langle  {\bar q} q \rangle$ itself
is systematically smaller then the ``standard'' value of 
$- (220 \mathrm{MeV})^3$.
It should be noted in this context that both the QCD sum rules value from ref.\cite{Nar2}
and the quenched lattice values from ref.\cite{Schierholz} refer to $\mu =1\mathrm{GeV}^2$,
whereas the instanton liquid model \cite{weiss} has a renormalization point which is given 
by the instanton size and therefore somewhat smaller $\approx (600\mathrm{MeV})^2$.
 
\begin{table}[ht]
\caption{$\langle  {\bar q} q \rangle$
and $ g_s \langle  {\bar q} \sigma G q \rangle$
at $\mu = 1\mathrm{GeV}^2$ for 
$g_s^2 D(s)  = (4\pi^2 d )\, \frac{\chi^2}{s^2 + \Delta}$,
$d = \frac{12}{27}$.
In all cases $f_\pi = 87 \mathrm{MeV}$.
Also displayed are the chiral low energy coefficients
$L_i,i=1,3,5,8$.}
\begin{tabular}{|c|c||c|c||c|c|c|c|}
\hline
$\Delta $ & $\chi $ & 
$-\langle \bar{q}q\rangle $  &
$-  g_s \langle  {\bar q} \sigma G q \rangle$ &
$L_1 $ & $L_3 $ & $L_5 $ & $L_8 $ 
\\ \hline
$[\mathrm{GeV}^4]$ & $ [\mathrm{GeV}]$ & $[\mathrm{MeV}^3]$ & $ [\mathrm{MeV}^5]$ & 
$*10^3$ & $*10^3$ & $*10^3$ & $*10^3$ 
\\ \hline \hline
$1*10^{-1}$ & $1.77$ & $(183)^3$ & $(459)^5$ & $0.79$ & $-3.76$ & $2.38$ & $1.03$ \\ \hline
$1*10^{-2}$ & $1.33$ & $(178)^3$ & $(457)^5$ & $0.80$ & $-4.01$ & $1.88$ & $0.93$ \\ \hline
$1*10^{-4}$ & $0.95$ & $(175)^3$ & $(456)^5$ & $0.81$ & $-4.23$ & $1.48$ & $0.88$ \\ \hline
$1*10^{-6}$ & $0.77$ & $(171)^3$ & $(452)^5$ & $0.83$ & $-4.41$ & $1.19$ & $0.84$ \\ \hline
\end{tabular}
\vspace{1cm}
\label{tab1}
\end{table}

\begin{table}[ht]
\caption{The same as in tab.\protect{\ref{tab1}} for
$g_s^2 D(s) = 3 \pi^2 \, \frac{\chi^2}{\Delta^2} \, \e^{-\frac{s}{\Delta} }$.}
\begin{tabular}{|c|c||c|c||c|c|c|c|}
\hline
$\Delta $ & $\chi $ & 
$-\langle \bar{q}q\rangle $  &
$-  g_s \langle  {\bar q} \sigma G q \rangle$ &
$L_1 $ & $L_3 $ & $L_5 $ & $L_8 $ 
\\ \hline
$[\mathrm{GeV}^2]$ & $ [\mathrm{GeV}]$ & $[\mathrm{MeV}^3]$ & $ [\mathrm{MeV}^5]$ & 
$*10^3$ & $*10^3$ & $*10^3$ & $*10^3$ 
\\ \hline \hline
$0.200$ & $1.55$ & $(183)^3$ & $(448)^5$ & $0.81$ & $-4.07$ & $1.56$ & $0.80$ \\ \hline
$0.020$ & $1.39$ & $(167)^3$ & $(431)^5$ & $0.83$ & $-4.43$ & $0.96$ & $0.81$ \\ \hline
$0.002$ & $1.23$ & $(151)^3$ & $(395)^5$ & $0.85$ & $-4.46$ & $0.82$ & $0.93$ \\ \hline
\end{tabular}
\vspace{1cm}
\label{tab2}
\end{table}

\begin{table}[ht]
\caption{The same as in tab.\protect{\ref{tab1}} for
$g_s^2 D(s) =  3 \pi^2 \,\frac{\chi^2}{\Delta^2} \, \e^{-\frac{s}{\Delta} } \,
+ \, \frac{4\pi^2 d}{s \ln [s/\Lambda^2 + \e ] }$, $d = \frac{12}{27}$,
$\Lambda = 200 \mathrm{MeV}$.}
\begin{tabular}{|c|c||c|c||c|c|c|c|}
\hline
$\Delta $ & $\chi $ & 
$-\langle \bar{q}q\rangle $  &
$-  g_s \langle  {\bar q} \sigma G q \rangle$ &
$L_1 $ & $L_3 $ & $L_5 $ & $L_8 $ 
\\ \hline
$[\mathrm{GeV}^2]$ & $ [\mathrm{GeV}]$ & $[\mathrm{MeV}^3]$ & $ [\mathrm{MeV}^5]$ & 
$*10^3$ & $*10^3$ & $*10^3$ & $*10^3$ 
\\ \hline \hline
$0.200$ & $1.65$ & $(173)^3$ & $(458)^5$ & $0.81$ & $-4.03$ & $1.66$ & $0.83$ \\ \hline
$0.020$ & $1.55$ & $(160)^3$ & $(448)^5$ & $0.82$ & $-4.39$ & $1.13$ & $0.84$ \\ \hline
$0.002$ & $1.45$ & $(151)^3$ & $(432)^5$ & $0.85$ & $-4.43$ & $0.97$ & $0.88$ \\ \hline
\end{tabular}
\vspace{1cm}
\label{tab3}
\end{table}

\begin{table}[t]
\caption{ }
\begin{tabular}{|c||c|c|}
\hline
 & $-\langle \bar{q}q\rangle^{\frac{1}{3}}\, [\mathrm{MeV}] $ &
$-  \langle g_s  {\bar q} \sigma G q \rangle^{\frac{1}{5}} \, [\mathrm{MeV}]$
\\ \hline \hline
this paper                                    & $150-180$ & $400-460$  \\ \hline
QCD sum rules \protect{\cite{Nar2}}           & $210-230$ & $375-395$  \\ \hline
quenched lattice \protect{\cite{Schierholz}}  & $225$     & $402-429$  \\ \hline
instanton liquid model \protect{\cite{weiss}} & $272$     & $490$      \\ \hline  
\end{tabular}
\vspace{1cm}
\label{tab4}
\end{table}

To summarize: 
We have considered a truncation of QCD
which gives rise to an effective quark-quark
interaction whose kernel is given by the model 
gluon 2-point function.
This gluon 2-point function is chosen to reproduce $f_\pi$ 
and it also reproduces the physics of the low energy 
mesonic sector.
We have shown how to obtain vacuum matrix elements
for any quark an gluon operator at the mean field level,
which is consistent with the large $N_C$ limit if the 
quark-quark interaction is fixed.
In particular we have found that any gluon operator reduces to 
an interacting multi-quark operator, whose vacuum expectation 
value can then be calculated in the mean field vacuum.
We have applied this technique to calculate the mixed 
quark-gluon condensate and found a value compatible 
with the ones of various other nonperturbative approaches.

\begin{ack}
I would like to thank L.S.Kisslinger for numerous 
very useful discussions and
comments.
This work has been supported by the NSF grant \# PHY-9319641.
\end{ack}

\end{document}